\documentclass[preprint]{revtex4}
\usepackage{amsmath}

\usepackage{mathrsfs}
\usepackage{comment}
\usepackage{amssymb}
\usepackage{graphicx}
\usepackage{subfigure}
\usepackage[colorlinks,
            linkcolor=blue,
            anchorcolor=blue,
            citecolor=blue]{hyperref}
\newtheorem{theorem}{Theorem}
\newtheorem{corollary}{Corollary}

\def\ra{\rangle}
\def\la{\langle}
\allowdisplaybreaks[4]

\begin{document}

\title{A note on uncertainty relations of metric-adjusted skew information}
\author{Qing-Hua Zhang$^{1,2,}$\footnotemark[1]}
\author{Jing-Feng Wu$^{1}$}
\author{Xiaoyu Ma$^{3}$}
\author{Shao-Ming Fei$^{1,4,}$\footnotemark[1]}

\affiliation{$^1$School of Mathematical Sciences, Capital Normal University, 100048
Beijing, China\\
$^2$Laboratoire de Physique Th\'eorique, CNRS, UPS, Universit\'e de Toulouse (UMR 5152), 31062 Toulouse, France\\
$^3$Zhongtai Securities Institute for Financial Studies, Shandong University, 250100 Jinan, China \\
$^4$Max-Planck-Institute for Mathematics in the Sciences, 04103 Leipzig, Germany}

\renewcommand{\thefootnote}{\fnsymbol{footnote}}

\footnotetext[1]{Corresponding authors. \\
\href{mailto:2190501022@cnu.edu.cn}{2190501022@cnu.edu.cn(Q. H. Zhang)}.\\
\href{mailto:feishm@cnu.edu.cn}{feishm@cnu.edu.cn(S. M. Fei)}.}
\bigskip

\bigskip
\begin{abstract}
The uncertainty principle is one of the fundamental features of quantum mechanics and plays a vital role in quantum information processing. We study uncertainty relations based on metric-adjusted skew information for finite quantum observables. Motivated by the paper [Physical Review A 104, 052414 (2021)], we establish tighter uncertainty relations in terms of different norm inequalities. Naturally, we generalize the method to uncertainty relations of metric-adjusted skew information for quantum channels and unitary operators. As both the Wigner-Yanase-Dyson skew information and the quantum Fisher information are the special cases of the metric-adjusted skew information corresponding to different Morozova-Chentsov functions, our results generalize some existing uncertainty relations. Detailed examples are given to illustrate the advantages of our methods.
\end{abstract}

\maketitle

%%%%%%%%%%%%%%%%%%%%%%%%%%%%%%%%%%%%%%%%%%%%%%%%%%%%%%%%%%%%%%%%%%%%%%%%
\section{Introduction}\label{sec1}

The uncertainty principle is one of most distinguished features of quantum mechanics, which shows the intrinsic differences between classical and quantum theory. Since Heisenberg first established the uncertainty principle of position and momentum \cite{Heisenberg1927}, uncertainty relations have been extensively studied. In 1929, Robertson proposed the famous uncertainty relation for any quantum state $|\psi\rangle$ and arbitrary two observables $A$ and $B$, $\Delta A\Delta B\geq \frac{1}{2}|\la \psi |[A,B]|\psi\ra|$, where $[A,B]=AB-BA$ is the commutator and $\Delta M=\sqrt{\la M^2\ra-\la M\ra^2}$ is the standard quantum deviation \cite{PhysRev.34.163}. With the development of quantum information theory, many kinds of uncertainty relations have been established, such as the ones in terms of entropy \cite{Maassen1988PhysRevLett.60.1103,Wu2009PhysRevA.79.022104,Coles2017RevModPhys.89.015002}, noise and disturbance \cite{buschPhysRevLett.111.160405}, skew information \cite{luoPhysRevLett.91.180403}, successive measurements \cite{DeutschPhysRevLett.50.631,DistlerPhysRevA.87.062112}, majorization techniques \cite{Pucha_a_2013}, etc. As one of the building blocks of quantum theory, uncertainty relations have many potential applications, including, but not limited to, quantum entanglement \cite{guhnePhysRevLett.92.117903,MacconePhysRevLett.113.260401,Zhang_2020A}, quantum coherence \cite{yuanPhysRevA.96.032313}, quantum steering \cite{SchneelochPhysRevA.87.062103}, quantum key distribution \cite{fuchsPhysRevA.53.2038}, and so on.

This paper focuses on quantum uncertainty relations characterized by the metric-adjusted skew information defined by Hansen \cite{Hansen9909}. The metric-adjusted skew information establishes a connection between the geometrical formulation of quantum statistics proposed by Chentsov and Morozova and the measures of quantum information introduced by Wigner and Yanase. When one takes particular Chentsov-Morozova functions, the metric-adjusted skew information redues to some common measures of quantum information such as the Wigner-Yanase skew information \cite{Wigner1963INFORMATION}, the Wigner-Yanase-Dyson skew information and quantum Fisher information \cite{luo2004wigner}. Recall that the Wigner-Yanase skew information of a state $\rho$ with respect to an observable $A$ is given by \cite{Wigner1963INFORMATION}
\begin{equation}
I_{\rho}(A)=-\frac{1}{2}{\rm Tr}([\sqrt{\rho},A]^2).
\end{equation}
Note that the skew information describes the non-commutativity between the square root of $\rho$ and the observable $A$, coinciding with the variance that describes the non-commutativity. Although the Wigner-Yanase skew information  is the same as the variance for pure states, but generally it is fundamentally different from the variance \cite{Luo2004On}. Later, Dyson generalized the skew information from the square root of $\rho$ to arbitrary $\alpha$ root, called the Wigner-Yanase-Dyson skew information,
\begin{equation}
I^{\alpha}_\rho(A):=-\frac{1}{2} {\rm Tr}(\left[\rho^{\alpha}, A\right]\left[\rho^{1-\alpha}, A\right]), \quad 0<\alpha<1.
\end{equation}
The celebrated convexity of $I^{\alpha}_\rho(H)$ with respect to the quantum state $\rho$ was successfully proved by Lieb \cite{LIEB1973267}. The quantum Fisher information is defined by
\begin{equation}
    I^F_\rho (A)=\frac{1}{4}{\rm Tr}(\rho L^2),
\end{equation}
where $L$, the symmetric logarithmic derivative, is a Hermitian operator satisfying
\begin{equation}
    i[\rho,A]=\frac{1}{2}(L\rho+\rho L).
\end{equation}

Recently, Cai shew that the sum uncertainty relations of Wigner-Yanase skew information also hold for metric-adjusted skew information, and presented a series of lower bounds given by the skew information of any prescribed size of the combinations \cite{Cai2021Sum}. Later, Ren $et\ al.$ proposed tighter uncertainty relations based on metric-adjusted skew information by using norm properties \cite{PhysRevA.104.052414}.

By using different operator norm inequalities we derive tighter sum uncertainty relations via metric-adjusted skew information for quantum observables, quantum channels and unitary operators. We point out that our new uncertainty relations can also be generalized to variance-based sum uncertainty relations. We start by introducing the definition of the metric-adjusted skew information in Sec.~\ref{sec2}. In Sec.~\ref{sec3}, we establish a tighter uncertainty relation of the metric-adjusted skew information for finite quantum observables. We generalize uncertainty relations to finite quantum channels and unitary operators in Sec.~\ref{sec4}. We summarize our results in Sec.~\ref{sec6}.

%%%%%%%%%%%%%%%%%%%%%%%%%%%%%%%%%%%%%%%%%%%%%%%%%%%%%%%%%%%%%%%%%%%%%%%%
\section{Metric-adjusted skew information}\label{sec2}

Let us recall the definition of metric-adjusted skew information. In Ref.~\cite{PETZ199681}, Petz introduced the symmetric monotone metric on the space of complex matrices by a positive semi-definite matrix of trace 1. Let $M_n(\mathbb{C})$ be the space of complex matrices of dimension $n$ and $\mathscr{M}_n(\mathbb{C})$ be the set of all positive semi-definite matrices of trace 1. Assuming $A,B\in M_n(\mathbb{C})$ and $\rho\in \mathscr{M}_n(\mathbb{C})$, the metric $K_\rho(\cdot,\cdot): M_n(\mathbb{C})\times M_n(\mathbb{C})\rightarrow \mathbb{C}$ satisfies the following conditions:
\begin{itemize}
    \item[(a)] $K_\rho(\cdot,\cdot)$ is sesquilinear;
    \item[(b)] $K_\rho(A,A)\geq 0$ and the equality holds if and only if $A=0$;
    \item[(c)] $K_\rho(\cdot,\cdot)$ is continuous on $\mathscr{M}_n(\mathbb{C})$ for every $\rho$;
\end{itemize}
Recall a linear map ${\rm T}: M_n(\mathbb{C})\rightarrow M_m(\mathbb{C})$ is defined to be stochastic if ${\rm T}(\mathscr{M}_n(\mathbb{C}))\subset \mathscr{M}_m(\mathbb{C})$ and ${\rm T}$ is completely positive.
\begin{itemize}
     \item[(d)] $K_{{\rm T}(\rho)}({\rm T}(A),{\rm T}(A))\leq K_\rho(A,A)$ for every stochastic map ${\rm T}: M_n(\mathbb{C})\rightarrow M_m(\mathbb{C})$.
\end{itemize}

A monotone metric has been given by Chentsov, Morozova and Petz \cite{PETZ199681},
\begin{equation}
    K_\rho(A,B)={\rm Tr}(A^\dagger c(L_\rho,R_\rho)B),
\end{equation}
where $c(L_\rho,R_\rho)$ is called the Morozova–Chentsov function with respect to the left and right multiplication operators $L_\rho$ and $R_\rho$. The Morozova–Chentsov function is of the form
\begin{equation}
    c(x,y)=\frac{1}{yf(xy^{-1})},\qquad x,\ y>0,
\end{equation}
where $f$ is a positive operator monotone function defined in the
positive half-axis satisfying the functional equation
\begin{equation}
f(t)=tf(t^{-1}),\qquad t>0.
\end{equation}

The metric-adjusted skew information given by the symmetric monotone metric is defined as \cite{Hansen9909,GIBILISCO20092225}:
\begin{equation}
\begin{aligned}
I_{\rho}^{c}(A) &=\frac{m(c)}{2} K_{\rho}^{c}(i[\rho, A], i[\rho, A]) \\
&=\frac{m(c)}{2} {\rm Tr} (i[\rho, A] c\left(L_{\rho}, R_{\rho}\right) i[\rho, A]),
\end{aligned}
\end{equation}
where $A$ is a self-adjoint operator and $m(c)=\lim_{t\rightarrow0}f(t)$.

If one considers
\begin{equation}
    f_\alpha=\alpha(1-\alpha)\frac{(1-t)^2}{(1-t^\alpha)(1-t^{1-\alpha})},\qquad t>0,
\end{equation}
with $f(0)=\alpha(1-\alpha)$, the corresponding Morozova-Chentsov function is
\begin{equation}\label{dysonmetric}
c_{\alpha}(x, y)=\frac{1}{\alpha(1-\alpha)} \frac{\left(x^{\alpha}-y^{\alpha}\right)\left(x^{1-\alpha}-y^{1-\alpha}\right)}{(x-y)^{2}},\qquad 0<\alpha<1.
\end{equation}
The metric-adjusted skew information of $c_\alpha$ becomes the Wigner-Yanase-Dyson skew information, that is,
\begin{equation}
\begin{aligned}
I_{\rho}^{c_{\alpha}}(A) &=\frac{\alpha(1-\alpha)}{2} \operatorname{Tr}\left\{i[\rho, A] c_{\alpha}\left(L_{\rho}, R_{\rho}\right) i[\rho, A]\right\} \\
&=-\frac{1}{2} \operatorname{Tr}\left[\rho^{\alpha}, A\right]\left[\rho^{1-\alpha}, A\right]=I_{\rho}^{{\alpha}}(A).
\end{aligned}
\end{equation}
Specially, if $\alpha=1/2$, $I_{\rho}^{c_{\alpha}}(A)$ becomes the Wigner-Yanase skew information.

If one takes
\begin{equation}
    f_F(t)=\frac{1+t}{2},\qquad t>0,
\end{equation}
the corresponding Morozova-Chentsov function is
\begin{equation}
    c_F(x,y)=\frac{2}{x+y},
\end{equation}
The metric-adjusted skew information of $c_F$ turns out to be
\begin{equation}
 I_{\rho}^{c_F}(A)=\frac{1}{4}{\rm Tr}(\rho L^2) =I^F_\rho(A)  .
\end{equation}
Obviously, the metric-adjusted skew information gives a uniform formula for the Wigner-Yanase-Dyson skew information and the quantum Fisher information.

%%%%%%%%%%%%%%%%%%%%%%%%%%%%%%%%%%%%%%%%%%%%%%%%%%%%%%%%%%%%%%%%%%%%%%%%
\section{Uncertainty relations for quantum observables}\label{sec3}
We now provide stronger sum uncertainty inequalities based on the
metric adjusted skew information for $N$ finite observables.

\begin{theorem} \label{th1}
For $N$ arbitrary finite observables $A_1, A_2, \dots, A_N$, the following  sum uncertainty relation holds:
\begin{equation}\label{th1eq1}
\begin{aligned}
\sum_{i=1}^N I^c_{\rho} (A_i)\geq \max_{x\in\{0,1\}}
&\frac{1}{2N-2}\Bigg\{
\frac{2}{N(N-1)} \Big [\sum_{1\leq i<j\leq N} \sqrt{I^c_{\rho}(A_i+(-1)^x A_j)}\Big ]^2\\
&+\sum_{1\leq i<j\leq N} I^c_{\rho} (A_i+(-1)^{x+1} A_j) \Bigg\}.
\end{aligned}
\end{equation}
\end{theorem}

{\sf [Proof]}
Using the following identity for any complex matrix $a_i$,
\begin{equation*}
(2N-2)\sum_{i=1}^{N} \| a_i\|^2 = \sum_{1\leq i<j \leq N} \| a_i+a_j \|^2 + \sum_{1\leq i<j \leq N} \| a_i-a_j \|^2,
\end{equation*}
where  $\|\cdot \|$ stands for the operator norm of a matrix,
and the Cauchy-Schwarz inequalities,
\begin{equation*}
\sum_{1\leq i<j \leq N} \| a_i + (-1)^x a_j \|^2 \geq \frac{2}{N(N-1)}
(\sum_{1\leq i<j \leq N} \| a_i +(-1)^x  a_j \|)^2,
\end{equation*}
for $x\in\{0,1\}$. Then one has
\begin{equation}\label{th1pf1}
\sum_{i=1}^{N} \| a_i\|^2 \geq \frac{1}{2N-2}\Big [\frac{2}{N(N-1)}(\sum_{1\leq i<j \leq N} \| a_i  +(-1)^x a_j \|)^2 + \sum_{1\leq i<j \leq N} \| a_i  +(-1)^{x+1} a_j \|^2\Big ].
\end{equation}
Namely,
\begin{equation}
\begin{aligned}
\sum_{i=1}^{N} K_{\rho}^{c}\left(i\left[\rho, A_{i}\right], i\left[\rho, A_{i}\right]\right) \geqslant
&\frac{1}{2N-2} \Bigg\{\sum_{1 \leqslant i<j \leqslant N} K_{\rho}^{c}\left(i\left[\rho, A_{i}+(-1)^{x+1} A_{j}\right], i\left[\rho, A_{i}+(-1)^{x+1} A_{j}\right]\right)\\
+&\frac{2}{N(N-1)}\Big [\sum_{1 \leqslant i<j \leqslant N} \sqrt{K_{\rho}^{c}\left(i\left[\rho, A_{i} +(-1)^x A_{j}\right], i\left[\rho, A_{i}+(-1)^x A_{j}\right]\right)}\Big ]^{2}
\Bigg\}.
\end{aligned}
\end{equation}
$\Box$

In particular, if we take the Morozova-Chentsov function to be $c_\alpha$, then the metric adjusted skew information becomes the Wigner-Yanase-Dyson skew information. In this case, we have the following corollary.
\begin{corollary}\label{cor1}
For $N$ arbitrary finite observables $A_1, A_2, \dots, A_N$, the following  sum uncertainty relation based on Wigner-Yanase-Dyson skew information holds:
\begin{equation}
\begin{aligned}
\sum_{i=1}^N I^{\alpha}_{\rho} (A_i)\geq \max_{x\in\{0,1\}}
&\frac{1}{2N-2}\Bigg\{
\frac{2}{N(N-1)} \Big [\sum_{1\leq i<j\leq N} \sqrt{I^{\alpha}_{\rho}(A_i+(-1)^x A_j)}\Big ]^2\\
&+\sum_{1\leq i<j\leq N} I^{\alpha}_{\rho} (A_i+(-1)^{x+1} A_j) \Bigg\}.
\end{aligned}
\end{equation}
\end{corollary}

When $\alpha=1/2$, the corollary gives rise to the uncertainty relation of Wigner-Yanase skew information, and covers the result in Ref.~\cite{zhang2021tighter}. If we take $c_F$ as the Morozova-Chentsov function, we have the following uncertainty relation of quantum Fisher information.

\begin{corollary} \label{cor2}
For $N$ arbitrary finite observables $A_1, A_2, \dots, A_N$, the following  sum uncertainty relation based on Fisher information holds:
\begin{equation}
\begin{aligned}
\sum_{i=1}^N I^{F}_{\rho} (A_i)\geq \max_{x\in\{0,1\}}
&\frac{1}{2N-2}\Bigg\{
\frac{2}{N(N-1)} \Big [\sum_{1\leq i<j\leq N} \sqrt{I^{F}_{\rho}(A_i+(-1)^{x} A_j)}\Big ]^2\\
&+\sum_{1\leq i<j\leq N} I^{F}_{\rho} (A_i+(-1)^{x+1} A_j) \Bigg\}.
\end{aligned}
\end{equation}
\end{corollary}

For pure quantum states, the sum uncertainty relation based on metric-adjusted skew information is equivalent to the sum variance-based uncertainty relation taking $c_\alpha$ or $c_F$ as the corresponding Morozova-Chentsov functions. In fact, for general mixed states, we also have the similar variance-based uncertainty relations for observables, which can be proved by using the inequality (\ref{th1pf1}).

\begin{corollary}\label{cor3}
For $N$ arbitrary finite observables $A_1, A_2, \dots, A_N$, the following  sum uncertainty relation based on quantum variance holds:
\begin{equation}\label{cor320}
\begin{aligned}
\sum_{i=1}\Delta^2_{\rho} (A_i)\geq \max_{x\in\{0,1\}}
&\frac{1}{2N-2}\Bigg\{\frac{2}{N(N-1)} \Big [\sum_{1\leq i<j\leq N} {\Delta_{\rho}(A_i+(-1)^{x} A_j)}\Big ]^2\\
&+\sum_{1\leq i<j\leq N} \Delta^2_{\rho} (A_i+(-1)^{x+1} A_j) \Bigg\},
\end{aligned}
\end{equation}
\end{corollary}
where $\Delta^2_{\rho} (A_i)=\|(A-\la A\ra)\sqrt{\rho}\|_F^2=:\|a_i\|_F^2$, $\|\cdot\|_F$ is the Frobenius norm.

In Ref.~\cite{PhysRevA.104.052414}, based on different norm inequality Ren $et\ al.$ presented another uncertainty relation for $N$ quantum observables,
\begin{equation}\label{renth1}
\begin{aligned}
\sum_{i=1}^{N} I_{\rho}^{c}\left(A_{i}\right) \geqslant \frac{1}{N} I_{\rho}^{c}(\sum_{i=1}^{N} A_{i}) +\frac{2}{N^{2}(N-1)}\Big [\sum_{1 \leqslant i<j \leqslant N} \sqrt{I_{\rho}^{c}\left(A_{i}-A_{j}\right)}\Big ]^{2} .
\end{aligned}
\end{equation}

In a similar fashion to the corollary \ref{cor3}, one may deduce the following result from (\ref{renth1}),
\begin{corollary}\label{cor4}
For $N$ arbitrary finite observables $A_1, A_2, \dots, A_N$, one has the following  sum uncertainty relation based on quantum variance:
\begin{equation}\label{cor422}
\sum_{i=1}\Delta^2_{\rho} (A_i)\geq
\frac{1}{N} \Delta_{\rho}^{2}(\sum_{i=1}^{N} A_{i}) +\frac{2}{N^{2}(N-1)}\Big [\sum_{1 \leqslant i<j \leqslant N} {\Delta_{\rho}\left(A_{i}-A_{j}\right)} \Big ]^{2}.
\end{equation}
\end{corollary}

Our Theorem \ref{th1} is tighter than the uncertainty relation (\ref{renth1}) for all quantum states and observables. The uncertainty relation (\ref{renth1}) is based on the inequality of matrix norm:
\begin{equation}\label{renpf1}
\sum_{i=1}^N\| a_i\|^2 \geqslant \frac{1}{N}\|\sum_{i=1}^N   a_i\|^2+\frac{2}{N^2(N-1)} (\sum_{1 \leqslant i<j \leqslant N}\|  a_i-  a_j\| )^2.
\end{equation}
To show that Theorem \ref{th1} is tighter than (\ref{renth1}) is equivalent to prove that
\begin{equation}\label{compareiqs}
\begin{aligned}
&\frac{1}{2N-2}\Big [\frac{2}{N(N-1)}(\sum_{1\leq i<j \leq N} \| a_i  - a_j \|)^2 + \sum_{1\leq i<j \leq N} \| a_i  + a_j \|^2\Big ] \\
&\geq
 \frac{1}{N}\|\sum_{i=1}^N   a_i\|^2+\frac{2}{N^2(N-1)} (\sum_{1 \leqslant i<j \leqslant N}\|  a_i-  a_j\| )^2.
\end{aligned}
\end{equation}
By using the following equality of matrix norm,
$$
\frac{1}{2N-2}\sum_{1\leq i<j \leq N} \| a_i  + a_j \|^2 = \frac{1}{N}\|\sum_{i=1}^N   a_i\|^2+\frac{N-2}{N(2N-2)} \sum_{1 \leqslant i<j \leqslant N}\|  a_i-  a_j\| ^2,
$$
the inequality (\ref{compareiqs}) becomes the Cauchy-Schwarz inequality,
\begin{equation*}
\sum_{1\leq i<j \leq N} \| a_i - a_j \|^2 \geq \frac{2}{N(N-1)}
(\sum_{1\leq i<j \leq N} \| a_i -  a_j \|)^2.
\end{equation*}
Therefore, our Theorem \ref{th1} is strictly tighter than the uncertainty relation (\ref{renth1}). By taking the maximum in (\ref{th1eq1}) for $x\in \{0,1\}$, the Theorem \ref{th1} would be much tighter. Below we consider a concrete example for illustration.

\begin{figure}[tbp]
 \centering
 \subfigure[]
 {
 \label{fig:subfig:a} %% label for first subfigure
 \includegraphics[width=7.8cm]{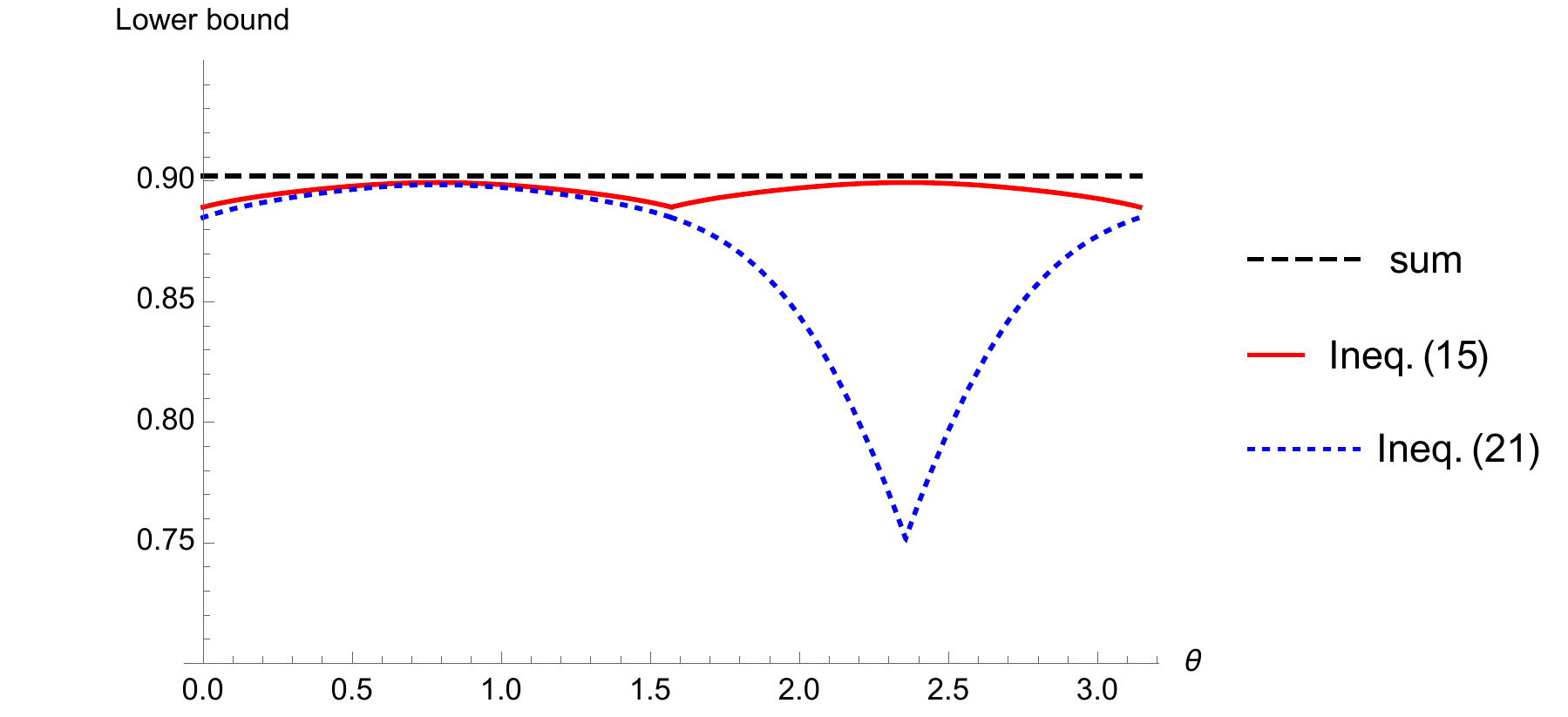}
 }
 \subfigure[]
 {
 \label{fig:subfig:b} %% label for second subfigure
 \includegraphics[width=7.8cm]{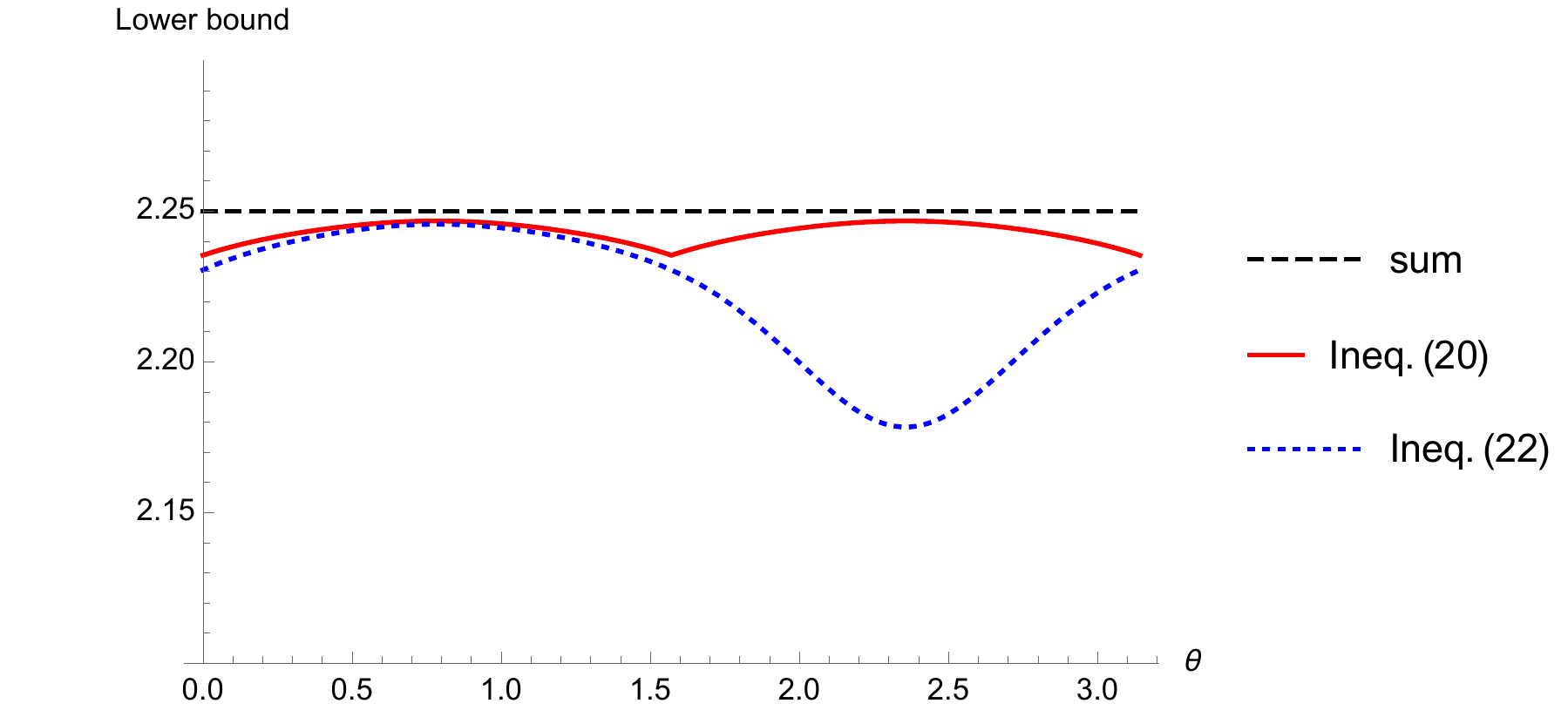}}
 \caption{(\textbf{a}) The black (dashed) line represents ${\rm sum}=I^{{1/3}}_{\rho} (\sigma_x)+I^{{1/3}}_{\rho} (\sigma_y)+I^{{1/3}}_{\rho} (\sigma_z)$. The red (full) curve and the blue (dotted) curve respectively denote the right hand of (\ref{th1eq1}) and (\ref{renth1}). Obviously, the bound given by the red curve is much tighter than the one corresponding to the blue curve. (\textbf{b}) The black (dashed) line represents the sum $\Delta^2_{\rho} (\sigma_x)+\Delta^2_{\rho} (\sigma_y)+\Delta^2_{\rho} (\sigma_z))$. The red (full) curve and the blue (dotted) curve respectively denote the lower bounds (\ref{cor320}) and (\ref{cor422}).
}
 \label{figex1}
 \end{figure}

{\emph{Example 1}}  Let us consider the mixed state $\rho=\frac{1}{2}(I_2+\vec{r}\cdot\vec{\sigma})$, where $\vec{r}=(\frac{\sqrt{3}}{2}\cos\theta,\frac{\sqrt{3}}{2}\sin\theta,0)$, the components of the vector$\vec{\sigma}=(\sigma_x,\sigma_y,\sigma_z)$ are the standard Pauli matrices, $I_2$ is the $2\times 2$ identity matrix.  We take $\alpha=1/3$ and the Pauli matrices $\sigma_x$, $\sigma_y$ and $\sigma_z$ to be the observables $A_1$, $A_2$ and $A_3$. The results are shown in Fig. \ref{figex1}.

%%%%%%%%%%%%%%%%%%%%%%%%%%%%%%%%%%%%%%%%%%%%%%%%%%%%%%%%%%%%%%%%%%%%%%%%
\section{Uncertainty relations for quantum channels}\label{sec4}
Let $\Phi$ be a quantum channel with Kraus representation, $\Phi(\rho)=\sum_{i=1}^{n}K_i\rho K_i^\dagger$.
In Ref.~\cite{PhysRevA.98.012113}, Luo and Sun proposed the Wigner-Yanase skew information of a channel with respect to a state $\rho$, which is well-defined due to the independence on the choice of the Kraus representations of the channel. The metric-adjusted skew information can be defined similarly \cite{Cai2021Sum}:
\begin{equation}
I^c_{\rho}(\Phi)=\sum_{i=1}^{n}I^c_{\rho}(K_i),
\end{equation}
where
\begin{equation}
\begin{aligned}
I_{\rho}^{c}(K_i) &=\frac{m(c)}{2} K_{\rho}^{c}(i[\rho, K_i], i[\rho, K_i]) \\
&=\frac{m(c)}{2} \operatorname{Tr} ((i[\rho, K_i])^\dagger c\left(L_{\rho}, R_{\rho}\right) i[\rho, K_i]).
\end{aligned}
\end{equation}
If we take $c=c_\alpha$, the metric-adjusted skew information for quantum channel becomes
\begin{equation}
I_{\rho}^{c_{\alpha}}(\Phi):=\sum_{i} I_{\rho}^{c_{\alpha}}\left(K_{i}\right)=-\frac{1}{2} \sum_{i} \operatorname{Tr}[\rho^{\alpha}, K_{i}^{\dagger}]\left[\rho^{1-\alpha}, K_{i}\right],
\end{equation}
Which corresponds to Wigner-Yanase-Dyson skew information. If $\alpha=1/2$, the metric-adjusted skew information above recovers the one given in \cite{PhysRevA.98.012113}.

The metric-adjusted skew information for a quantum channel $\Phi$ can be rewritten as
\begin{equation}
\begin{aligned}
I_{\rho}^{c}(\Phi) &=\frac{m(c)}{2} \operatorname{Tr} (\sum_i (i[\rho, K_i])^\dagger c\left(L_{\rho}, R_{\rho}\right) i[\rho, K_i]) \\
&=\frac{m(c)}{2}{\rm Tr}(a^\dagger C\left(L_{\rho}, R_{\rho}\right) a),
\end{aligned}
\end{equation}
where $a^\dagger=((i[\rho,K_1])^\dagger, (i[\rho,K_2])^\dagger, \dots, (i[\rho,K_n])^\dagger)$ and $C\left(L_{\rho}, R_{\rho}\right)=I_n \otimes c\left(L_{\rho}, R_{\rho}\right)$. $I^c_{\rho}(\Phi)$ characterizes the intrinsic features of both the quantum state and the quantum channel \cite{PhysRevA.98.012113}. For arbitrary $N$ quantum channels, we have the following conclusion.

\begin{theorem}\label{th2}
Let $\Phi_1,\Phi_2,\dots,\Phi_N$ be $N$ quantum channels with Kraus representations $\Phi_s(\rho)=\sum_{i=1}^{n}K_i^s\rho (K_i^s)^\dagger$, $s=1,2,...,N$. The following uncertainty relation holds:
\begin{equation}
\sum_{s=1}^{N}I^c_{\rho}(\Phi_s) \geq \max \{{\rm LB1},{\rm  LB2}, {\rm LB3}\},
\end{equation}
where
\begin{equation}
\begin{aligned}
{\rm LB1}= \max_{\pi_s,\pi_t \in S_n}
&\frac{1}{N-2}\Bigg \{\sum_{1\leq s<t\leq N}\sum_{i=1}^n I^c_{\rho}(K_{\pi_s(i)}^s+K_{\pi_t(i)}^t) \\
&-\frac{1}{(N-1)^2}\Bigg [\sum_{1\leq s<t\leq N}\sqrt{\sum_{i=1}^n I^c_{\rho}(K_{\pi_s(i)}^s+K_{\pi_t(i)}^t)} \Bigg ]^2\Bigg \},
\end{aligned}
\end{equation}
\begin{equation}
\begin{aligned}
{\rm LB2}= \max_{\pi_s,\pi_t \in S_n}
&\frac{1}{N}\Bigg \{\sum_{i=1}^n I^c_{\rho}(\sum_{s=1}^N K_{\pi_s(i)}^s) \\
&+ \frac{2}{N(N-1)}\Bigg [\sum_{1\leq s<t\leq N}\sqrt{\sum_{i=1}^n I^c_{\rho}(K_{\pi_s(i)}^s-K_{\pi_t(i)}^t)} \Bigg ]^2\Bigg \},
\end{aligned}
\end{equation}
\begin{equation}
\begin{aligned}
{\rm LB3}= \max_{x\in\{0,1\}}\max_{\pi_s,\pi_t \in S_n}
&\frac{1}{2N-2}\Bigg \{\sum_{1\leq s<t\leq N}\sum_{i=1}^n I^c_{\rho}(K_{\pi_s(i)}^s+(-1)^{x} K_{\pi_t(i)}^t) \\
&+ \frac{2}{N(N-1)}\Bigg [\sum_{1\leq s<t\leq N}\sqrt{\sum_{i=1}^n I^c_{\rho}(K_{\pi_s(i)}^s+(-1)^{x+1} K_{\pi_t(i)}^t)} \Bigg ]^2\Bigg \},
\end{aligned}
\end{equation}
$\pi_s,\pi_t \in S_n$ are arbitrary $n$-element permutations.
\end{theorem}

{\sf [Proof]} Similar to the proof of Theorem \ref{th1}, we need to use three inequalities for the operator norm $\|\cdot\|$,
\begin{equation*}
\begin{aligned}
\sum_{s=1}^N\|a_s\|^2 \geq &\frac{1}{N-2} [\sum_{1\leq s<t\leq N}\|a_s+a_t\|^2-\frac{1}{(N-1)^2}(\sum_{1\leq s<t\leq N}\|a_s+a_t\|)^2],            \\
\sum_{s=1}^N \|a_s\|^2 \geq &\frac{1}{N}[\|\sum_{s=1}^N a_s\|^2+\frac{2}{N(N-1)}(\sum_{1\leq s<t \leq N}\|a_s-a_t\|)^2],                    \\
\sum_{s=1}^{N} \| a_s\|^2 \geq & \frac{1}{2N-2}[\frac{2}{N(N-1)}(\sum_{1\leq s<t \leq N} \| a_s +(-1)^x a_t \|)^2 + \sum_{1\leq s<t \leq N} \| a_s+(-1)^{x+1} a_t \|^2],
\end{aligned}
\end{equation*}
where $x\in \{0,1\}$. Using these three inequalities, one can straightforwardly derive the lower bounds LB1, LB2 and LB3 in (\ref{th1eq0}). $\Box$

In Ref.~\cite{PhysRevA.104.052414}, Ren $et\ al.$ established tighter uncertainty relations of metric-adjusted skew information for $N$ quantum channels shown as following:
\begin{equation}\label{renth2}
\begin{aligned}
\sum_{s=1}^{N} I_{\rho}^{c}\left(\Phi_{s}\right) \geq
&\max _{\pi_{s}, \pi_{t} \in S_{n}} \frac{1}{N-2}\sum_{i=1}^{n}\Bigg\{\sum_{1 \leq s<t \leq N}  I_{\rho}^{c}\left(K_{\pi_{s}(i)}^{s}+K_{\pi_{t}(i)}^{t}\right)       \\
&-\frac{1}{(N-1)^{2}}\Big [\sum_{1 \leq s<t \leq N} \sqrt{I_{\rho}^{c}(K_{\pi_{s}(i)}^{s}+K_{\pi_{t}(i)}^{t})}\Big ]^{2}\Bigg\}
\end{aligned}
\end{equation}
and
\begin{equation}\label{renth3}
\begin{aligned}
\sum_{s=1}^{N} I_{\rho}^{c}\left(\Phi_{s}\right) \geq
&\max _{\pi_{s}, \pi_{t} \in S_{n}} \frac{1}{N}\sum_{i=1}^{n}\Bigg\{ I^c_{\rho}(\sum_{s=1}^N K_{\pi_s(i)}^s)       \\
&+\frac{2}{N(N-1)}\Big [\sum_{1 \leq s<t \leq N} \sqrt{I_{\rho}^{c}(K_{\pi_{s}(i)}^{s} - K_{\pi_{t}(i)}^{t})}\Big ]^{2}\Bigg\}.
\end{aligned}
\end{equation}

Let us compare the uncertainty relations (\ref{renth2}) and (\ref{renth3}) with our theorem \ref{th2} by an example.

\begin{figure}[tbp]
 \centering
 \subfigure[]
 {
 \label{subfigex2:a} %% label for first subfigure
 \includegraphics[width=7.8cm]{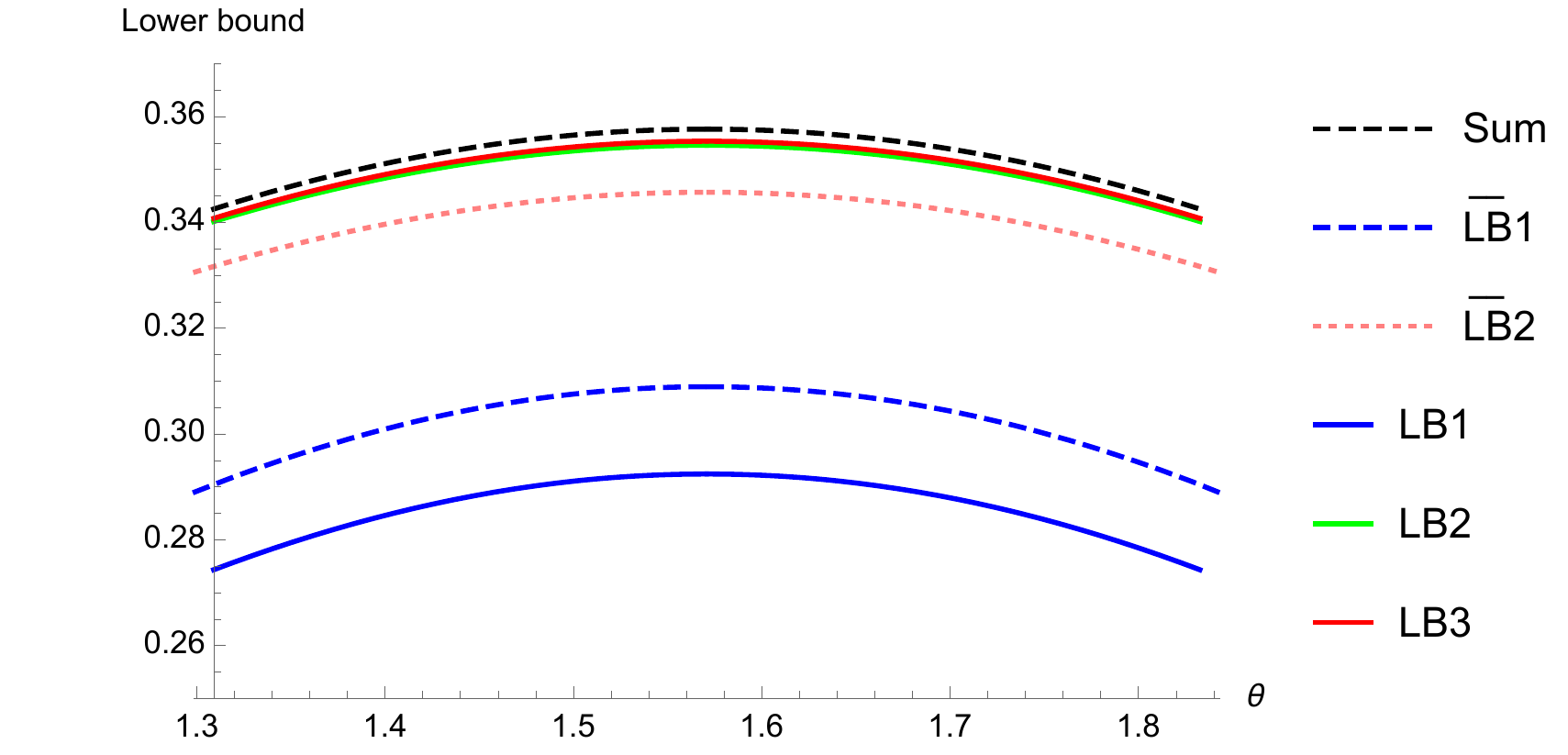}
 }
 \subfigure[]
 {
 \label{subfigex2:b} %% label for second subfigure
 \includegraphics[width=7.8cm]{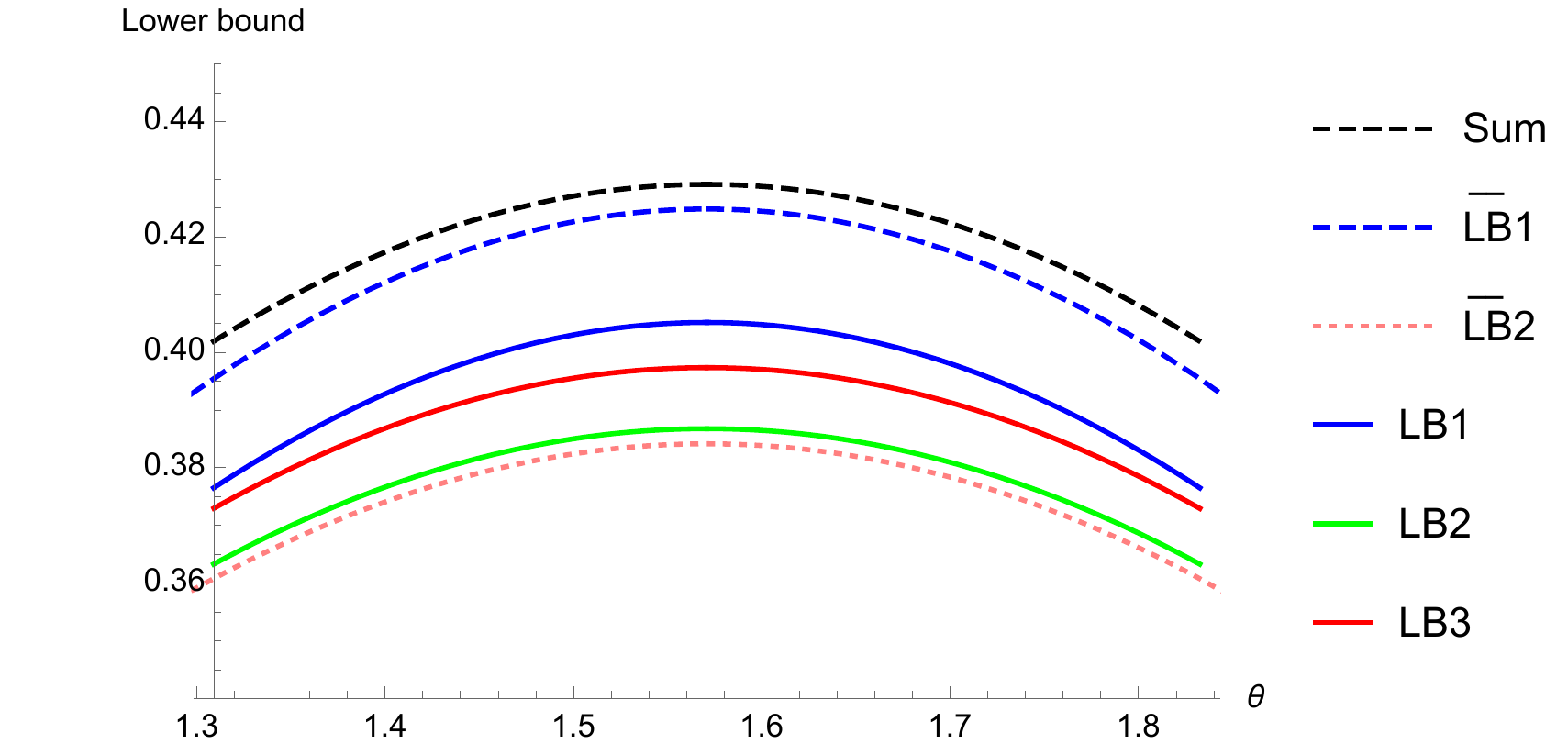}
 }
  \subfigure[]
  {
 \label{subfigex2:c} %% label for first subfigure
 \includegraphics[width=7.8cm]{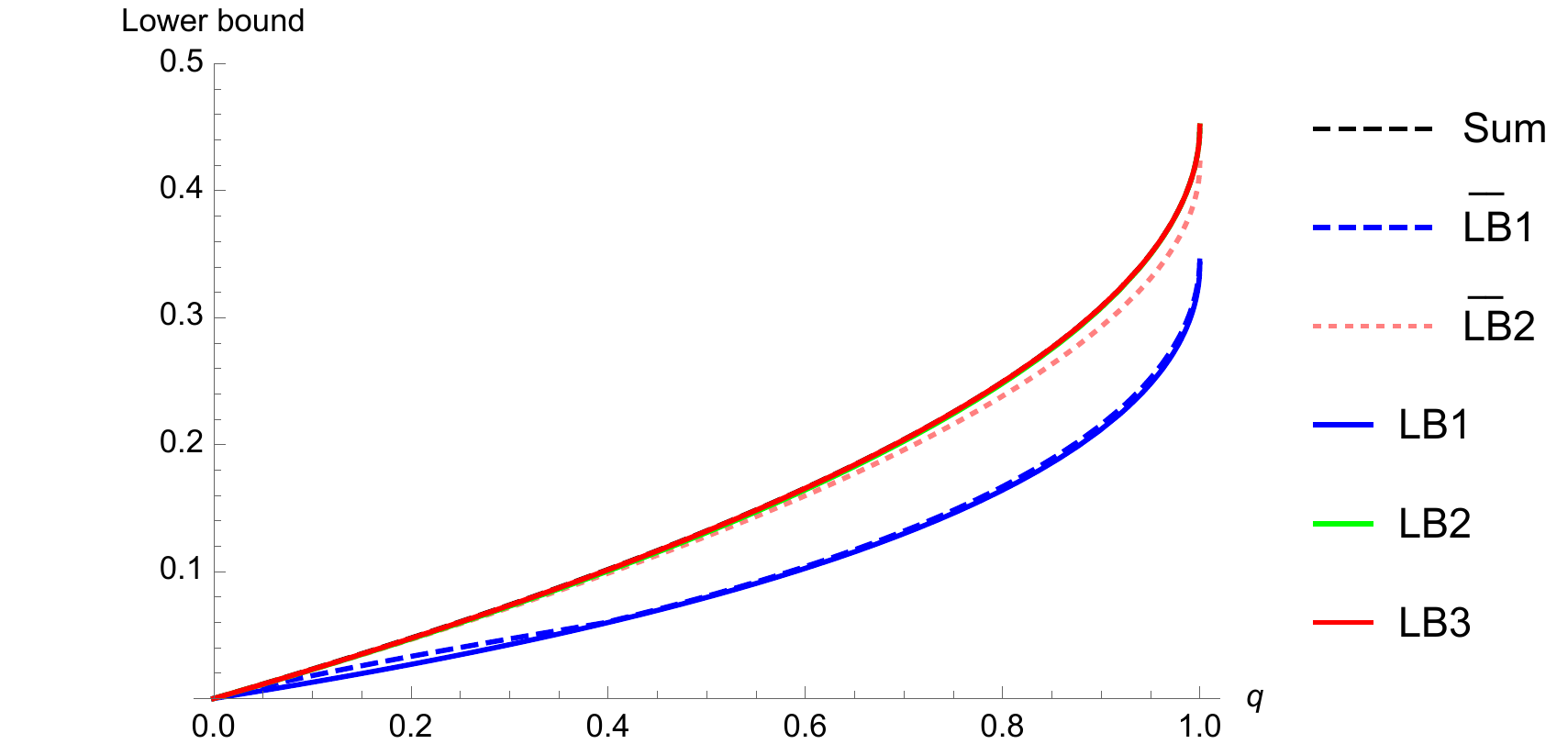}
 }
  \subfigure[]
  {
 \label{subfigex2:d} %% label for first subfigure
 \includegraphics[width=7.8cm]{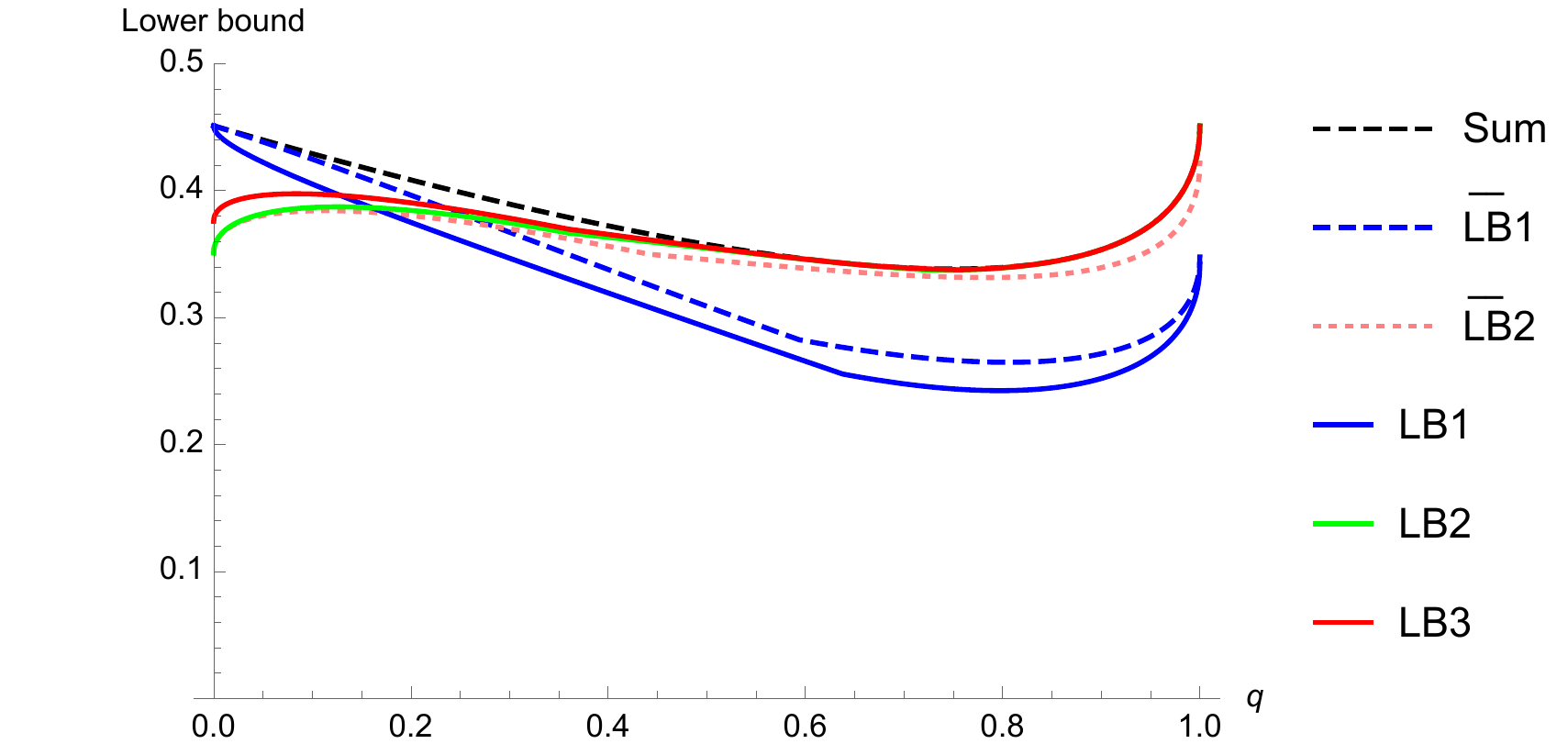}
 }
 \caption{The black (dashed) curve represents ${\rm Sum}=I^{1/3}_{\rho}(\phi)+I^{1/3}_{\rho}(\epsilon)+I^{1/3}_{\rho}(\Lambda)$. The red curve, the green curve and the blue curve respectively represent the lower bounds LB3, LB2 and LB1 in Theorem \ref{th2}. The pink (dotted) curve and blue (dashed) curve represent the lower bounds of (\ref{renth2}) and (\ref{renth3}), respectively. (\textbf{a}) $q=0.5$. One can see that our lower bounds {\rm LB2} and {\rm LB3} are tighter than ones in (\ref{renth2}) and (\ref{renth3}). (\textbf{b}) $q=0.1$. LB1 is tighter than LB2 and LB3. (\textbf{c}) $\theta=0$. LB2 and LB3 are tighter than ${\rm\overline{LB}1}$ and ${\rm\overline{LB}2}$ for all $q$. (\textbf{d}) $\theta=\pi/2$. LB2 and LB3 are the tighter ones for most choices of $q$.}
 \label{figex2}
 \end{figure}

{\emph{Example 2} Let us consider the mixed state $\rho=({I_2+\vec{r}\cdot\vec{\sigma}})/{2}$
defined by the Bloch vector $\vec{r}=(\frac{\sqrt{3}}{2}\cos\theta,\frac{\sqrt{3}}{2}\sin\theta,0)$.
We respectively consider three quantum channels: the amplitude damping channel $\epsilon$,
$$\epsilon(\rho)=\sum_{i=1}^2A_i\rho (A_i)^\dagger,\quad A_1=|0\ra\la0|+\sqrt{1-q}|1\ra\la1|,\quad A_2=\sqrt{q}|0\ra\la1|
,$$
the phase damping channel $\phi$,
$$
\phi(\rho)=\sum_{i=1}^2B_i\rho (B_i)^\dagger,\quad B_1=|0\ra\la0|+\sqrt{1-q}|1\ra\la1|,\quad B_2=\sqrt{q}|1\ra\la1|
$$
and the bit flip channel $\Lambda$,
$$
\Lambda(\rho)=\sum_{i=1}^2C_i\rho (C_i)^\dagger,\quad C_1=\sqrt{q}|0\ra\la0|+\sqrt{q}|1\ra\la1|,\quad C_2=\sqrt{1-q}(|0\ra\la1|+|1\ra\la0|)
$$
with $0\leq q<1$.
We use the Wigner-Yanase-Dyson skew information to compare the lower bounds with $\alpha=1/3$. For convenience, we denote ${\rm\overline{LB}1}$ and ${\rm\overline{LB}2}$ the value of the right hands of (\ref{renth2}) and (\ref{renth3}), respectively.

Our results are based on the norm inequalities. Nevertheless, there is no fixed relationship among the lower bounds of these norm inequalities. Hence, it is not guaranteed that the Theorem 2 is strictly better than the result of Ref.~\cite{PhysRevA.104.052414}. Fig. \ref{figex2} shows that our results can give tighter lower bounds than those of (\ref{renth2}) and (\ref{renth3}) from Ref.~\cite{PhysRevA.104.052414} in certain cases.

%%%%%%%%%%%%%%%%%%%%%%%%%%%%%%%%%%%%%%%%%%%%%%%%%%%%%%%%%%%%%%%%%%%%%%%%
%\section{Uncertainty relations for unitary operators}\label{sec5}
Naturally, we would like to consider uncertainty relations for unitary operators, which govern the evolution of a closed quantum system \cite{nielsen2002quantum}. Consider a unitary transformation $U(\rho)=U\rho U^\dagger$. The metric-adjusted skew information for the unitary operator can be similarly defined by
\begin{equation}
\begin{aligned}
  I_{\rho}^{c}(U) =&\frac{m(c)}{2} K_{\rho}^{c}(i[\rho, U], i[\rho, U])  \\
  =&\frac{m(c)}{2} {\rm Tr}((i[\rho, U])^\dagger c(L_\rho,R_\rho)i[\rho, U]).
\end{aligned}
\end{equation}
Similarly to the uncertainty relations for quantum observables, we have the following conclusion, whose proof is similar to one for Theorem \ref{th2}.

\begin{corollary}\label{th3}
The following  uncertainty relation of metric-adjusted skew information holds
for $N$ unitary operators $U_1, U_2,\dots, U_N$,
\begin{equation}\label{th1eq0}
\sum_{s=1}^{N}I^c_{\rho}(U_s) \geq {\rm max }\{{\rm Lb1},{\rm  Lb2}, {\rm Lb3}\},
\end{equation}
where
\begin{equation}
\begin{aligned}\label{th3eq1}
{\rm Lb1}:=
\frac{1}{N-2}\Bigg \{\sum_{1\leq s<t\leq N} I^c_{\rho}(U_s+U_t)
-\frac{1}{(N-1)^2}\Bigg [\sum_{1\leq s<t\leq N}\sqrt{ I^c_{\rho}(U_s+U_t)} \Bigg ]^2\Bigg \},
\end{aligned}
\end{equation}
\begin{equation}\label{th3eq2}
\begin{aligned}
{\rm Lb2}:=
\frac{1}{N}\Bigg \{I^c_{\rho}(\sum_{s=1}^N U_s)
+ \frac{2}{N(N-1)}\Bigg [\sum_{1\leq s<t\leq N}\sqrt{I^c_{\rho}(U_s-U_t)} \Bigg ]^2\Bigg \},
\end{aligned}
\end{equation}
\begin{equation}\label{th3eq3}
\begin{aligned}
{\rm Lb3}:=\max_{x\in\{0,1\}}
&\frac{1}{2N-2}\Bigg \{\sum_{1\leq s<t\leq N} I^c_{\rho}(U_s+(-1)^{x} U_t)\\
&+ \frac{2}{N(N-1)}\Bigg [\sum_{1\leq s<t\leq N}\sqrt{ I^c_{\rho}(U_s+(-1)^{x+1} U_t)} \Bigg ]^2\Bigg \}.
\end{aligned}
\end{equation}
\end{corollary}

In Ref.~\cite{Zhang_2021}, the authors presented uncertainty relations based on generalized Skew information for quantum unitary channels. In fact, our Corollary \ref{th3} covers the main results of Ref.~\cite{Zhang_2021}. We take an example to illustrate these unitary uncertainty relations.

{\emph{Example 3}}
 Let us consider the mixed state $\rho=\frac{1}{2}(I_2+\vec{r}\cdot\vec{\sigma})$ with $\vec{r}=(\frac{1}{\sqrt{2}}\cos\theta,\frac{1}{\sqrt{2}}\sin\theta,0)$.
Consider three unitary operators generated by standard Pauli matrices,
\begin{equation*}
%\begin{gathered}
U_1=e^{\frac{i\pi \sigma_x}{8}}=
\begin{pmatrix} \cos\frac{\pi}{8} & i \sin\frac{\pi}{8} \\ i \sin\frac{\pi}{8} &\cos\frac{\pi}{8} \end{pmatrix},~~
U_2=e^{\frac{i\pi \sigma_y}{8}}=
\begin{pmatrix} \cos\frac{\pi}{8}  & \sin\frac{\pi}{8} \\ -\sin\frac{\pi}{8}  & \cos\frac{\pi}{8}  \end{pmatrix},~~
U_3=e^{\frac{i\pi \sigma_z}{8}}=
\begin{pmatrix} e^{i \frac{\pi}{8}} & 0 \\ 0&e^{-i\frac{\pi}{8}} \end{pmatrix},
%\end{gathered}
\end{equation*}
which respectively correspond to the Bloch sphere rotations of $-\pi/4$ about $x$ axis, $y$ axis and $z$ axis. Using Wigner-Yanase-Dyson skew information with $\alpha=1/4$, the comparison among the lower bounds of Corollary \ref{th3} is shown in Fig. \ref{figex3}.
\begin{figure}[t]
 \centering
 \includegraphics[width=15cm]{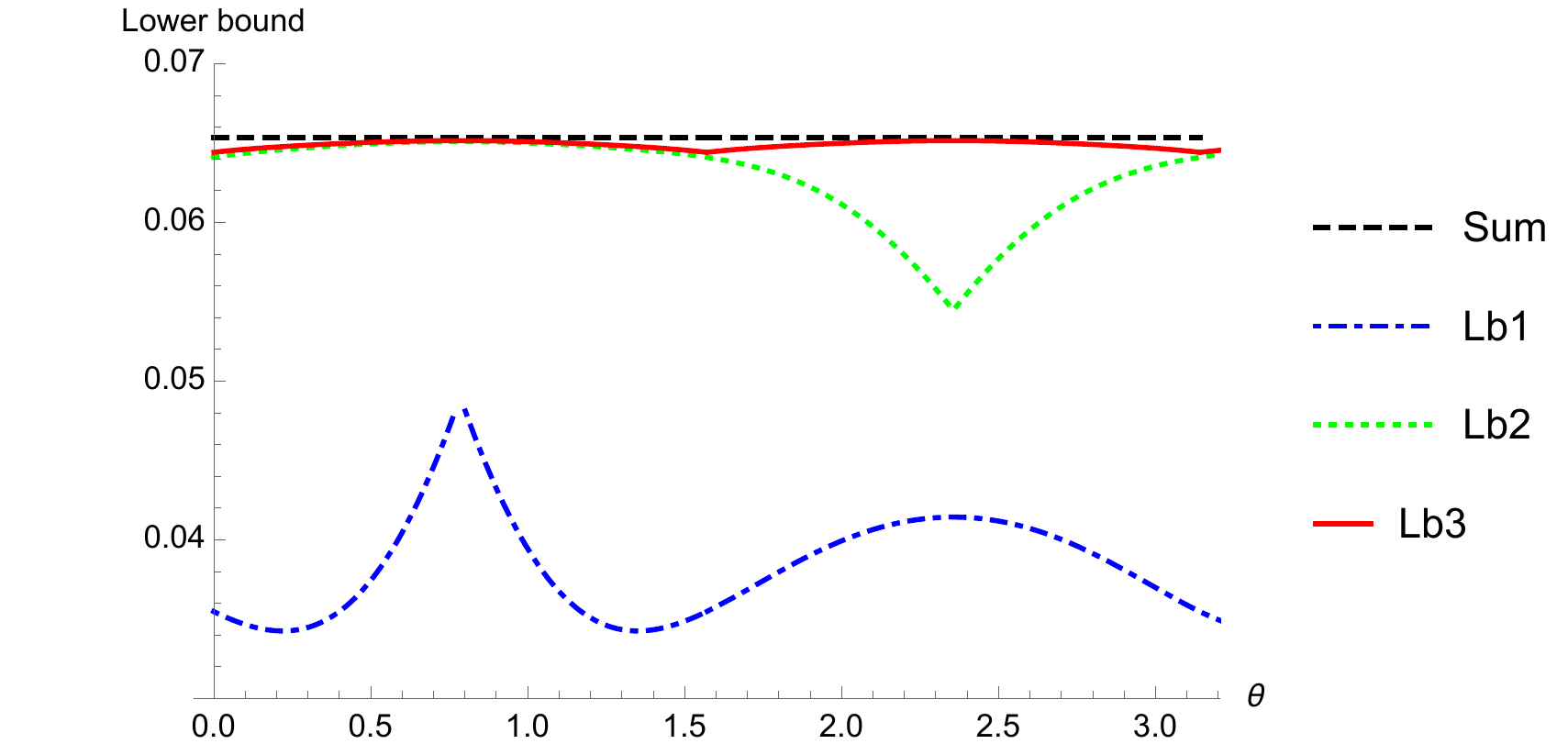}
 \caption{The black (dashed) line is the ${\rm Sum}=I^{1/4}_{\rho}(U_1)+I^{1/4}_{\rho}(U_2)+I^{1/4}_{\rho}(U_3)$. The red (full) curve Lb1, green (dotted) curve Lb2 and blue (dot-dashed) curve Lb3 represent the right sides of (\ref{th3eq1}), (\ref{th3eq2}) and  (\ref{th3eq3}), respectively.}
  \label{figex3}
 \end{figure}

%%%%%%%%%%%%%%%%%%%%%%%%%%%%%%%%%%%%%%%%%%%%%%%%%%%%%%%%%%%%%%%%%%%%%%%%
\section{Conclusion}\label{sec6}
We have studied uncertainty relations for metric-adjusted skew information. We provided tighter uncertainty relations for quantum observables, quantum channels, and also for unitary operators. We have established a uniform formula to describe the uncertainty, which includes, but is not limited to, Wigner-Yanase skew information, Wigner-Yanase-Dyson skew information, and quantum Fisher information. We also put forward the variance-based uncertainty relations for quantum observables. The results of our work are a starting point for further investigations on uncertainty relations and their related implications and applications.

\bigskip

\noindent{\emph{Note added in revision}\,  After we submitted this paper to QIP on Jan. 19, 2022, we noted that the paper arXiv:2205.09286 was submitted to Arxiv on May, 2022 \cite{Li2022Tighter}, in which the authors  considered a generalized version of our results on sum uncertainty relations with regard to metric-adjusted skew information by means of the parameterized norm inequalities.

\bigskip

\noindent{\bf Acknowledgments}\, We are grateful to I. Nechita for his comments on the paper. This work is supported by NSFC (Grant Nos.~12075159,~12171044), Beijing Natural Science Foundation (Z190005), the Academician Innovation Platform of Hainan Province, and China Scholarship Council.

\noindent{\bf Data availability}\, Data sharing not applicable to this article as no data sets were generated or analyzed during the current study.

\noindent{\bf Author contributions}\,
The first author wrote the main manuscript text and all authors reviewed and edited the manuscript.

\noindent{\bf Competing interests}\,
The authors declare no competing interests.

\bibliographystyle{sn-mathphys}
\bibliography{metricrefs}
\end{document}